# Experimental realization of three-dimensional elastic phononic topological insulator

Shao-yong Huo[1], Jiu-jiu Chen[1]*, Hong-bo Huang[1], Yong-jian Wei[2], Zhu-hua Tan[2]*, Lu-yang Feng[1], and Xiao-ping Xie[1]

[1]State Key Laboratory of Advanced Design and Manufacturing for Vehicle Body, College of Mechanical and Vehicle Engineering, Hunan University, Changsha 410082, People's Republic of China.

[2]School of Mechanical Engineering, Hebei University of Technology, Tianjin 300401, People's Republic of China.

*Correspondence and requests for materials should be addressed to J.J.C. (email: jjchen@hnu.edu.cn) or to Z.H.T. (email: zhtan@hnu.edu.cn)

Three-dimensional (3D) elastic phononic topological insulator, featuring two-dimensional (2D) surface states, which support the high-efficient and robust elastic wave propagation without backscattering in all spatial dimensions, remains a challenge due to the nature of multiple polarized elastic modes and their complex hybridization in 3D. Here, a 3D elastic phononic topological insulator is designed and observed experimentally by emulating the quantum valley Hall effects. The spatial inversion of adjacent atoms gives rise to a valley topological phase to an "insulating" regime with complete 3D topological phononic bandgap. The 2D surface states protected by valley topology are unveiled numerically, which are confirmed experimentally to have a great robustness against the straight channel and sharp bends. Further engineering the elastic valley layer with appropriate interlayer coupling, we also demonstrate that layer pseudospin can be created in 3D elastic system which leads to 2D topological layer-dependent surface states and layer-selective transport. Our work will be a key step for the manipulation of elastic wave in 2D topological plane and the applications of 3D elastic topological-insulator-based devices with layer-selective functionality.

Motivated by the condensed-matter topological systems, significant progress has recently been made in understanding and realizing the classical analogues of 2D topological insulators (TIs) for photonic[1-8], sonic[9-25] and elastic waves[26-38]. In spite of intrinsic difference between spin electrons and spinless classical bosons, the excellent merits like backscattering inhibition, edge-state confined, and robustness for fabrication imperfections can be inherited by creating the "spinlike" degree of freedom in bosonic systems. The early approach to engineer topological phase in classical 2D systems relies on the breaking of time-reversal symmetry by introductions of the bianisotropy[7,40], magnetic-optical effects[41] for photonic crystals, and circulating fluids[9-11], rotating gyroscopes[25] and spatiotemporal modulation[12-14] for phononic crystals (PnCs). Later, benefiting from the artificial symmetry of lattices, a new class of topological states has been developed by breaking the spatial inversion symmetry, two most representatives of which are quantum spin Hall (QSH) TIs[17,19,22-24,28-30,33,37,38] and quantum valley Hall (QVH) TIs[15,16,20,21,31,34-36]. In the 2D system, owing to the advantages of easy fabrication and high-efficiency wave guiding, abundant 2D TIs are designed and realized based on different macroscopic scales, configurations and controllability platforms, in which the 1D edge states have exhibited wide application perspectives, such as robust waveguides[17], topological directional antennas[21], topological laser and programmable topological devices[22,35]. In addition to the 2D TIs, the research of topological states in 3D is receiving a growing attention because 3D TIs possess the 2D gapless or gapped topological surface states that allow us to manipulate the wave propagation in full spatial directions[41-47]. Despite significant efforts devoted in 3D photonic[41,42] and acoustic TIs[45-47], the realization of 3D elastic phononic TIs, especially on an experimental observation, remains challenging owing to the elastic multimode hybridization in 3D space and fabrication difficulties.

Elastic wave propagation in PnC has exhibited several crucial advantages compared with the conventional material in manipulating elastic wave energy, such as vibration isolation[48], energy harvesting[49], damage detection and polarization filtering[50]. Different from electromagnetic and acoustic waves, elastic waves in PnC contain richer elastic polarizations (two transverse modes and one longitudinal mode) and their waveform conversion between transverse and longitudinal waves happens pervasively, which makes the elastic waves complex and hard to control. On the other hand, fabrication imperfections and environmentally induced deformations in solid are fairly ubiquitous and hard to avoid, which usually cause the dramatical attenuation and large losses of elastic energy. Fortunately, TIs open a new avenue to manipulate elastic waves owing to the key characteristic: the robust defect-immune transport. The 3D elastic PnC TIs offer an opportunity to manipulate the elastic wave in all three spatial directions, which can be an immense potential in engineering applications for wave guiding with extremely low losses, signal processing[51] with high signal-to-noise ratio and nondestructive testing with strong sensitivity. Previous extensive studies have shown that 1D topological edge states can be realized in 2D elastic QSH and QVH TIs[28-39]. Whereas, for the 3D elastic TIs, its realization still remain elusive so far due to several restrictions, such as complicated elastic mode coupling, high-demanding computational and experimental platforms. In addition, although the Weyl Fermi arc[52] and nodal surface states[53] have been found in 3D elastic system, these ungapped bulk states can not be used to confine the elastic wave. Therefore, a 3D elastic TI featuring 2D surface states in a complete topological bandgap is extremely desirable.

In this paper, we present the first numerical simulation and experimental observation of topological valley- and layer-dependent backscattering-free transports in 3D elastic PnCs. Firstly, the valley surface states for 3D elastic waves along 2D domain walls are demonstrated and the robust valley transports in three spatial directions are experimentally verified. Then, by stacking the two elastic layers with an appropriate twist, the layer pseudospin is successfully introduced into in the 3D valley-Hall system, which gives rise to the 2D layer-polarized surface states. The layer-direction locking behavior is found at the 2D surface states between the two adjacent elastic layers, and it leads to the selective layer-filtered and layer-beam-split features only by switching the place of excitation source, which are further validated by experiment. This realization of 3D robust topological propagation for elastic waves will be a significant step for developing 3D fabrication technology and advanced functional devices with high productivity and lower power consumption.

**3D elastic metacrystal and 2D surface states.**

We construct a 3D elastic PnC by periodically stacking the circular rods and patterned plate with array of hexagonal holes along the $z$ axis. Figure 1a shows a schematic of the 3D lattice model. A unit cell outlined by the red rhombus with lattice constant of $a$ = 34mm is detailedly shown in Fig. 1b. The height of the unit cell and the thickness of perforated plate are $h_0$ = 15.3mm and $h_1$ = 6.3mm, respectively. The top-view schematic depicting the unit cell is shown in Fig. 1c. The patterned plate for the unit cell can be considered as a combination consisting of two nonequivalent hexagonal chunks with side-lengths, $L_A$ and $L_B$ respectively, connected by a narrow vein with width of $w$ = 2.8mm. The whole structure is made by aluminum material with Young's modulus $E$ = 68.9Gpa, density $\rho$ = 2700kg/m$^3$, and Poisson's ratio $v$ = 0.33. The symmetry and topological phase transition of 3D elastic PnCs are controlled by the parameter $\Delta L = L_A - L_B$. The finite element method (FEM) is employed to calculate the band structure for different $\Delta L$.

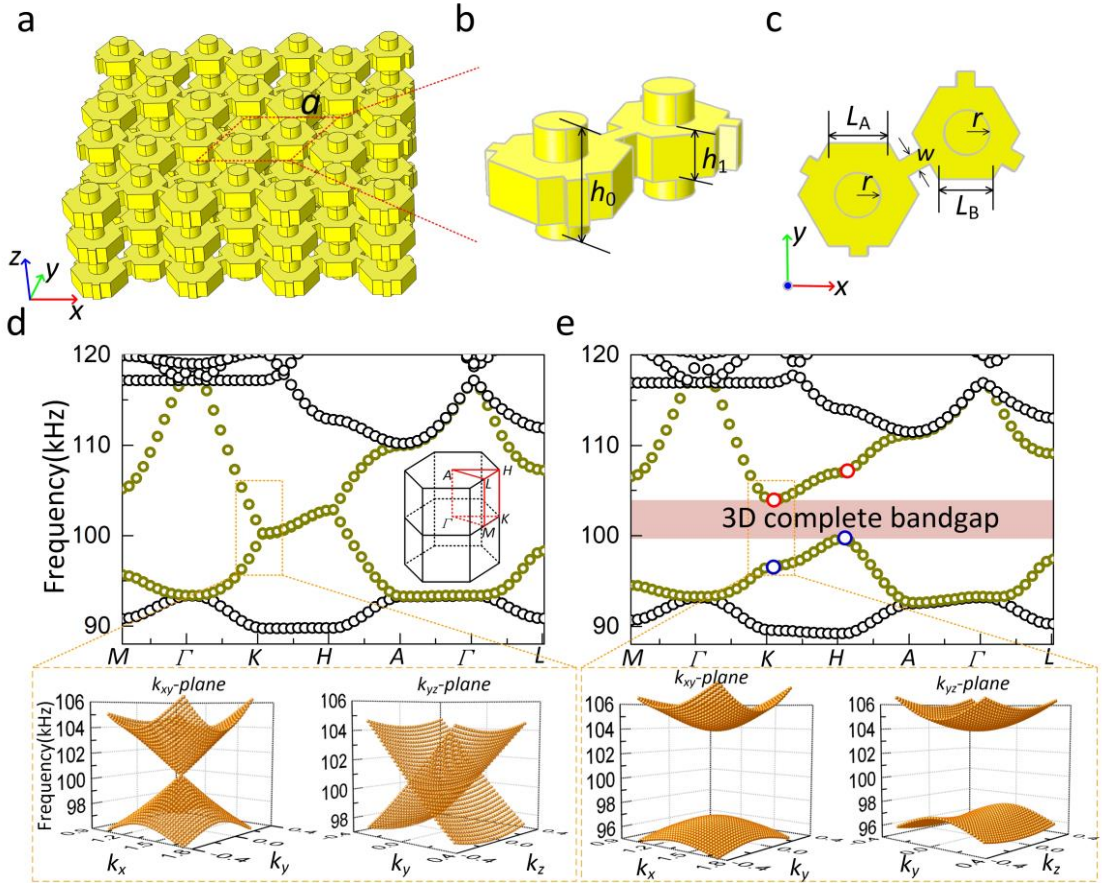

**Fig. 1 | 3D valley bulk states and phase inversion. a** schematic of the 3D model. **b, c** Side and top views of the unit cell with two nonequivalent chunks A and B. The patterned plates are linked by the circular rods ($r$ = 3.5mm) along the $z$ axis. **d, e** 3D band structures of bulk elastic wave with the identical chunks of $L_A = L_B$ = 9.5mm and that with the different chunks of $L_A$ = 9.8mm (9.2mm) and $L_B$ = 9.2mm (9.8mm), respectively. The orange dashed boxes show the bulk bands protected onto the $k_{xy}$-plane and

$k_{yz}$-plane near the K point. Along the KH direction, the bulk bands have the same properties.

When $\Delta L = 0$, the 3D elastic PnC is protected by spatial inversion symmetry and time-reversal symmetry. The bulk band structure is calculated as shown in Fig. 1d. A Dirac cone appears at the K point in half of the first Brillouin zone (BZ), which is formed by 14[th] and 15[th] bands predominately with transverse polarization (see Supplemental Note 1). The bulk bands near the K point projected on the $k_{xy}$-plane and $k_{yz}$-plane are plotted at the lower panel. We can see that the Dirac point stretches into a two-fold nodal line along the KH direction because of the $D_{6h}$ symmetry. Next, we consider two cases of $\Delta L = 0.6$mm and $\Delta L = -0.6$mm to break the in-plane inversion symmetry. The nodal line is split nearly parallel and a full 3D phononic band gap appears as presented in Fig. 1e. The eigenstates of two bands above and below the band gap are labeled by $p^-(q^+)$ and $q^+(p^-)$ for KH (K'H') nodal line, respectively. It is noted that for arbitrary $k_z$, two lattices of $\Delta L$ with opposite signs have distinct phases associated with non-zero Berry curvatures, which is connected by a band-inversion process when $\Delta L$ varies from positive to negative. The difference in the topological charge across the 2D interface can be quantized by topological invariants (valley Chern numbers) $|\Delta C_v| = 1$ (see Supplemental Note 2).

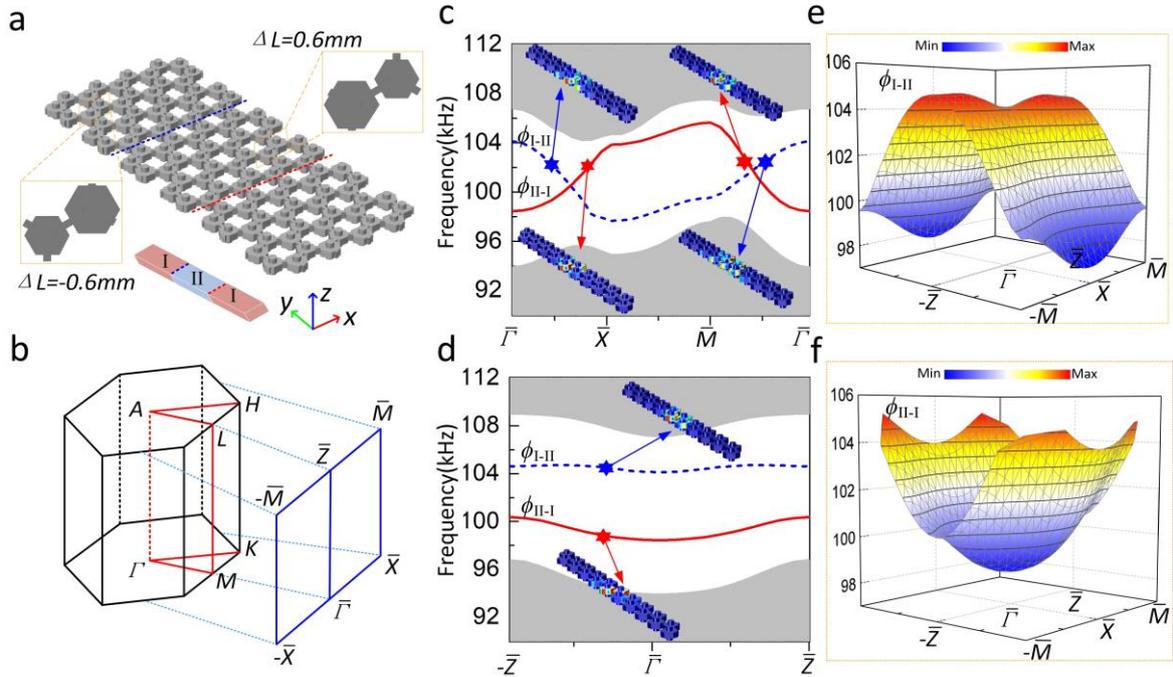

**Fig. 2 | 2D valley surface states. a** Schematic of two supercell interfaces constructed by two different PnCs with $\Delta L = 0.6$mm (denoted by type II) and $\Delta L = -0.6$mm (denoted by type I). **b** One-half of the

surface BZ projected onto the $k_{xz}$-plane. **c, d** Projected band dispersions for the interface separating two domains with different valley Hall phases and the total displacement fields of the denoted points on surface states. **e, f** The 2D valley surface states $\phi_{\text{I-II}}$ and $\phi_{\text{II-I}}$ at two different domain walls (types of I-II and II-I) for the whole projected surface BZ.

According to the bulk-boundary correspondence, the 2D surface states locate at interface of two 3D lattices with different valley-Hall phases because of the nonzero valley Chern numbers. To demonstrate it, we construct a zigzag-type domain wall between two distinct valley lattices with opposite $\Delta L$ as shown in Fig. 2a. There are two different interfaces marked by types I-II and II-I. Figure 2b shows the one-half of the surface BZ projected onto the $k_{xz}$ plane. The numerical dispersions for I-II and II-I type supercells are plotted in Figs. 2c and 2d, respectively. As predicted, two surface states $\phi_{\text{I-II}}$ and $\phi_{\text{II-I}}$ appear in the band gap, which correspond to two different interfaces, respectively. The inserted pictures also indicate that the displacement fields are well confined at the domain walls. For a full view, the 2D valley Hall surface states on the whole projected $k_{xz}$ plane are further plotted as shown in Figs. 2e and 2f.

**Experimental confirmation of robust 3D elastic wave transport.**

In contrast to the valley-projected 1D edge modes in 2D PnCs, the greatest superiority of the 2D surface states in 3D elastic system is that it can propagates in two dimensions, which allows us to test its robustness for different spatial directions by simulation and experiment. A fabricated sample of 3D elastic topological protected waveguide in a finite $8a \times 6a \times 10h_0$ is photographed as shown in Fig. 3a where the straight interface is designed separated by the two lattices ($\Delta L = 0.6$mm and $\Delta L = -0.6$mm) with different valley phases. The experimental configuration to measure the 2D surface states along the $\overline{\Gamma}\text{-}\overline{X}$ and $\overline{\Gamma}\text{-}\overline{Z}$ directions are displayed in Figs. 3b and 3c, respectively. In our experiment, a couple of standard shear wave transducers are employed as the emitting and receiving ends to measure the elastic wave propagation (see method). For the $\overline{\Gamma}\text{-}\overline{X}$ direction, the measured transmission spectra along for the surface states and bulk states are shown in Fig. 3d. We can see that the transmission of bulk states has an obvious drop from 98.2 kHz to 104.1 kHz due to the existence of 3D complete bandgap and the transmissions of surface states with straight domain wall are ~40 dB higher than it. The simulated displacement fields at the 101 kHz for the case of straight domain wall and without domain wall are shown in Fig. 3e. At the same time, the transmission spectra along the $\bar{\Gamma}$-$\bar{Z}$ direction are measured with a bulge band from 98.2 kHz to 100.5 kHz as shown in Fig. 3f and the corresponding simulated displacement fields are simulated (Fig. 3g). Last, to further test the robustness against the sharply curved path for the surface state, we construct a Z-shaped 3D domain wall and numerically simulate the displacement fields for several different out-of-plane wavenumber $k_z$ (see Supplemental Figure 4), in which the uniform displacement field intensity distributions indicate the wave propagates along the interface regardless of the bends (see Supplemental Note 3).

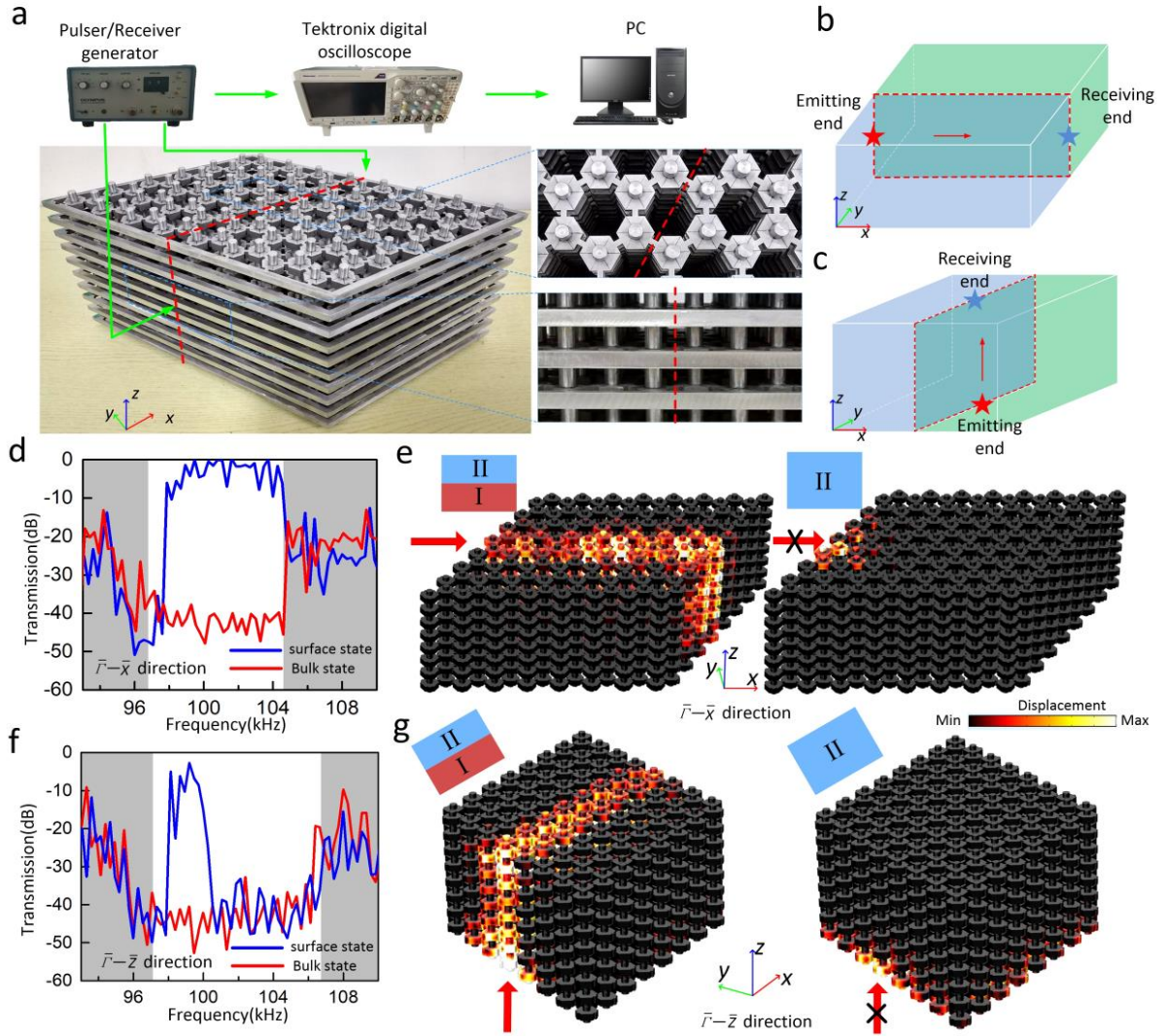

**Fig. 3 | Experimentally measured topological valley transport of the 3D elastic topological insulator.**
**a** Photograph of a fabricated 3D PnC sample (without the cover layers on the top, bottom, left and right sides) with the top-down and lateral views. **b, c** Experimental configurations to measure the surface
7

transmission spectra along the $\overline{\Gamma}$-$\overline{X}$ and $\overline{\Gamma}$-$\overline{Z}$ directions, respectively. The red dotted boxes mark the 2D domain walls along the *xz*-plane and *yz*-plane, respectively. **d** Measured transmission of the valley surface states along the $\overline{\Gamma}$-$\overline{X}$ directions for two cases: without the domain wall (red line) and with straight domain wall (blue line). **e** Displacement field distributions along the $\overline{\Gamma}$-$\overline{X}$ topological interface and trivial bulk region at the 101 kHz. **f** Measured transmission of the valley surface states along the $\overline{\Gamma}$-$\overline{Z}$ directions. **g** Displacement field distributions along the $\overline{\Gamma}$-$\overline{Z}$ topological interface and trivial bulk region at the 99.3 kHz.

**3D layered elastic topological insulators.**

It has been demonstrated that the intriguing layer-locking transport behavior can be generated in 2D valley-Hall acoustic and photonic insulators by introducing nonzero interlayer coupling[15,54,55]. Here we theoretically and experimentally demonstrate such propagation phenomena for the 2D layer-dependent surface states in 3D elastic waves system. We start the study from the band dispersion of the unit cell. As shown in Fig. 4a, the unit cell is doubled along the *z*-direction and the band structure is calculated by FEM. Similarly, two parameters $\Delta L_1$ and $\Delta L_2$ are used to characterize the symmetry of upper layer and lower layer PnCs, respectively. When $(\Delta L_1, \Delta L_2) = (0, 0)$, the nodal line is folded into a nodal ring covering the whole KH high-symmetric line. It is noted that when the $k_z$ is away from the K point to the H point, the ring degeneracy is formed, which resembles the case proposed in 2D bilayer sonic system[15]. When $k_z$ is changed to the H point ($k_z = \pi/h_0$), the ring degeneracy is narrowed to be a point forming a four-fold Dirac degeneracy, which can be seen from the 3D dispersion at the right panel of Fig. 4a. This nodal ring along the *z*-direction can be just right to be used construct the elastic layer-related pseudospins in 3D space. To break the spatial symmetry, we consider a glide symmetry design with $\Delta L_1 = -\Delta L_2$, which makes the two layers of PnCs twisted with respect to middle plane. When $(\Delta L_1, \Delta L_2)$ = (-0.6mm, 0.6mm), two layers are twisted that breaks the mirror symmetry and gapped the nodal ring to open a 3D full phononic band gap (Fig. 4b). Interestingly, when the height of rod becomes well-matched to height of the patterned plates, the interlayer coupling can be opened and the split eigenstates exhibit the distinctive polarization properties that the displacement fields are either confined to the upper layer or to the lower layer. As the $\Delta L_1$ and $\Delta L_2$ changes, the layer polarization undergoes a reversal process (see Supplemental Figure 5). Therefore, by integrating the layer information, the topological charge difference across a 3D PnC interface separating distinct layer-related valley Hall phases can be described as $\left|\Delta C_L^H\right|=2$ (see Supplemental Note 4). To explore the layer-related 2D surface states, we also calculate the supercell band structures for a 2D domain wall between two PnCs with $(\Delta L_1, \Delta L_2) = (\mp 0.6mm, \pm 0.6mm)$ (Figs. 4c, d). Excitedly, two 2D surface states carrying opposite group velocities occur in the omnidirectional gap, the displacement fields of which are locked at the interfaces of different layers. Moreover, these two surface states with layer distinct polarizations are crossed and formed two nearly parallel surface nodal lines (see Supplemental Figure 6). It is worth noting that the nearly-flat surface states can be obtained along the $k_z$ direction (Fig. 4d) due to the large distortion, which may produce significant applications in 3D elastic devices, such as notch filter and collimator.

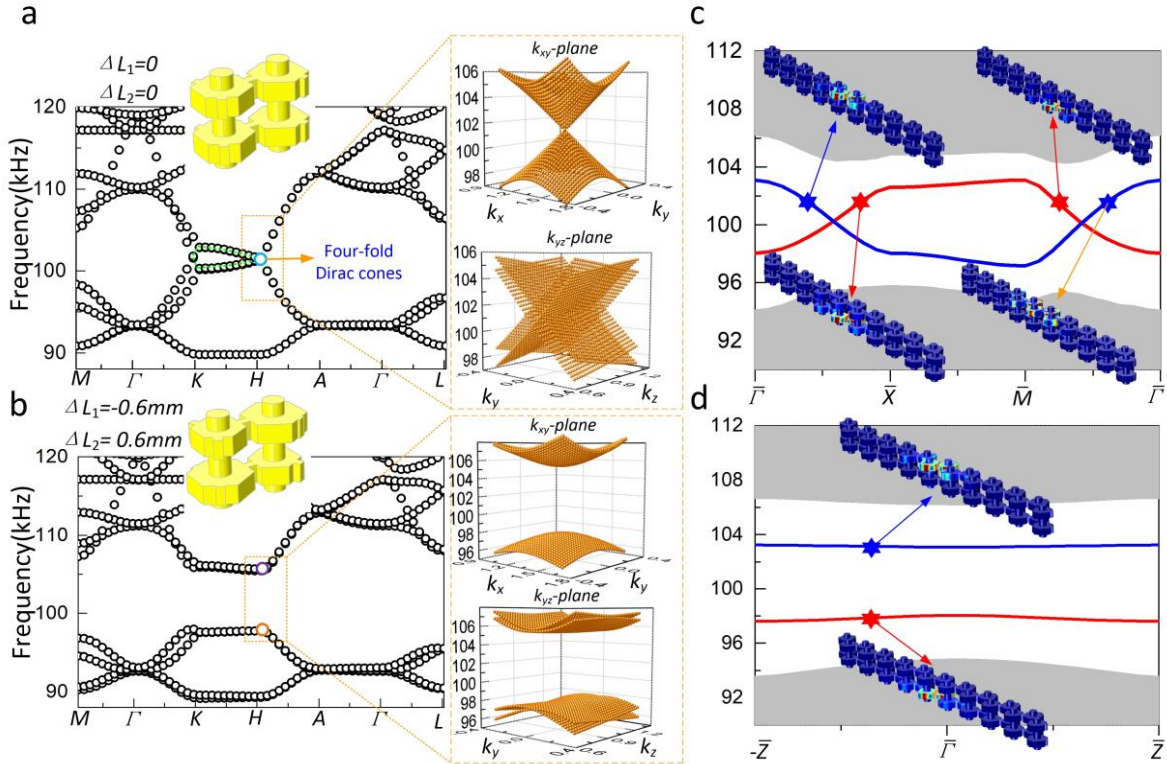

**Fig. 4 | Layer-dependent surface states for the 3D bilayer elastic PnCs. a** Band structure for the double-sized unit cell along the z-direction with $\Delta L=0$. **b** Band structure when two layer cells have opposite $\Delta L$ ($\Delta L_1$=0.6mm and $\Delta L_2$=-0.6mm). **c, d** Projected band dispersions for the interface separating

two domains with different layer-related valley Hall phases and the total displacement fields of the denoted points on surface states.

The layer-related topological surface states have been realized for the 3D elastic wave by engineering the interlayer symmetry. Then, we demonstrate the layer-selective robust transports in 3D elastic layer interface system by experiments. The 3D fabricated sample is shown in Supplemental Fig. 7, which is periodically stacked by the layer-1 (($\Delta L_1$, $\Delta L_2$) = (0.6mm, -0.6mm)) and the layer-2 (($\Delta L_1$, $\Delta L_2$) = (-0.6mm, 0.6mm)). In our experiments, a shear wave generator transducer is placed at the left entrance of the layer-1/layer-2, and a shear wave receiver transducer is placed at the right side of the layer-1/layer-2 to measure the transmission spectra (see Supplemental Note 6). For the $\overline{\Gamma}$-$\overline{X}$ direction, the measured transmission coefficients are displayed in Fig. 5a and present a large transport difference between two adjacent layers. For a detail, the displacement field distributions of two layers and that of fields on slices are plotted in Fig. 5b. We can see that when the excitation source is set at the layer-1, the surface states can pass through the layer-1 but be blocked at the layer-2. While the position of the excitation source is switched to the layer-2, the surface states can well pass through the layer-2 but be blocked at the layer-1, indicating that a 3D layer-selective transport of elastic wave can be easily achieved only by controlling the excitation of layer, which may inspire some interesting applications, such as elastic layer switchers and sensors. For the $\overline{\Gamma}$-$\overline{Z}$ direction, the measured transmission spectra and displacement field distributions are shown in Fig. 5c and 5d, respectively. However, it is worth noting that an almost single frequency 3D topological layer-polarized propagations with high-transmission along the $\overline{\Gamma}$-$\overline{Z}$ direction are realized, which exhibits a great potential in layer filter. In addition, by employing the property of elastic topological layer polarization, a 3D toplogical splitter with layer-selective function is numerically designed in Supplemental Fig. 8.

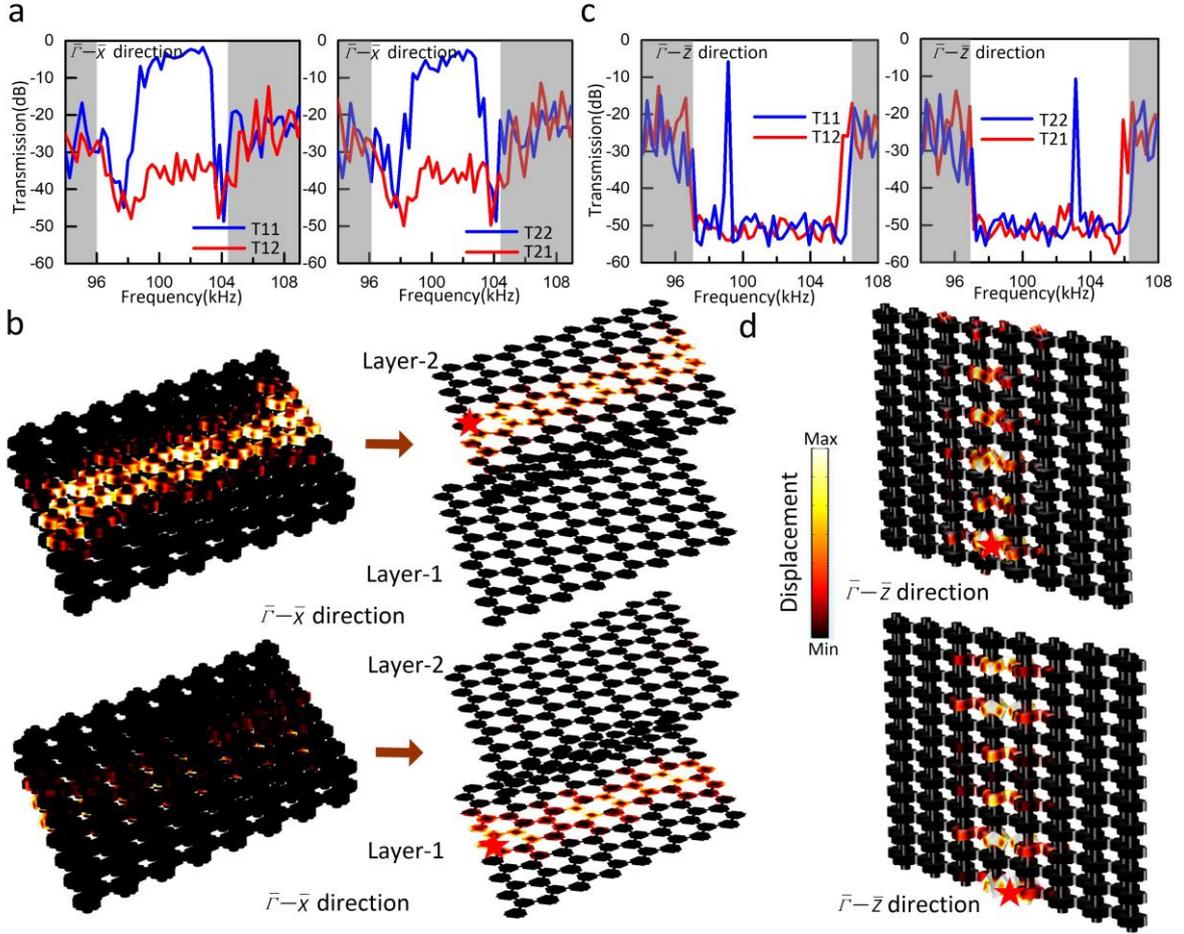

**Fig. 5 | 3D elastic layer-selective transports of 2D layer-polarized surface states. a, b** Experimentally measured transmission spectra of the layer-polarized surface states along the $\overline{\Gamma}$-$\overline{\text{X}}$ directions, $\overline{\Gamma}$-$\overline{\text{Z}}$ directions, respectively. The transmission representations T11, T12, T22 and T21 are used to illustrate the measured layer-polarized spectra where the first subscript refers to the emitting layer and the second subscript refers to the receiving layer. **c** Displacement field distributions with the point source excited from the layer-1 and layer-2 at 102 kHz along the $\overline{\Gamma}$-$\overline{\text{X}}$ direction. **d** Displacement field distributions with the point source excited from the layer-1 at 98.1 kHz and that from layer-2 at 103.1 kHz along the $\overline{\Gamma}$-$\overline{\text{Z}}$ direction.

**Discussion**

In summary, we have experimentally demonstrated the first implementation of 3D elastic valley and layered phononic TIs. The 3D valley topological transports with robustness for elastic wave are investigated experimentally in three spatial dimensions, which open the door to design the valley-related topological devices beyond 2D, such as 3D valley filter and splitter. The non-zero interlayer coupling has been introduced to produce the 2D layer-polarized surface states in 3D elastic system, which provides a new route to guide and control the elastic wave. The 3D elastic layer-related TIs exhibiting a novel layer-selective function paves the way for the design and applications in advance acoustic signal processing, elastic energy harvesting, integrated topological devices and phononic circuits in 3D geometries. Although our findings focus on the simple and macroscopic 3D elastic phononic system, the study may be extended to the microscopic scale and the other particle systems, such as heat, electricity and soft mechanical structure.

**Methods**

*Numerical simulations* - All numerical simulations are performed by using the commercial FEM software (COMSOL Multiphysics). The simulations are implemented in the 3D structural mechanics module, including the eigenfrequency study to calculate the 3D band structure and frequency domain study to calculate the displacement field distributions. The bulk band dispersions (Figs. 1d-e, 4a-b) are obtained by using one unit cell with Floquet periodic boundary conditions in all three directions. The surface state dispersions (Figs. 2c-f, 4c-d) were calculated by using a supercell ribbon structure, Floquet periodic boundary conditions are imposed to the periodic surface in two spatial directions. When performing the simulations of the displacement field distributions, the perfectly matched layers (PML) are set around the whole structure to prevent the leakage of energy.

*Experimental measurement* - In the experiment, the 3D PnC sample is stacked to 10 layers along the $z$ direction with every layer made of 24×24 unit cell in $xy$ plane. For the convenience of measurement, the sample on the upper, lower, left and right surfaces are seamlessly covered by a layer flat plate with thickness of 8mm, respectively. During the measurement, a couple of ultrasonic broadband shear wave transducers with a central frequency of 100 kHz and a diameter of 25 mm (OLYMPUS direct contact transducers type Videoscan No. V1548) are employed as the emitter and receiver. The emitter is excited with an ultrasonic emission source (OLYMPUS model 5073PR) to produce a short-duration pulse with large amplitude. The emitting transducer is placed at the center of the left (upper) side to launch a probing shear wave signal and the receiving transducer is placed at the center of right (bottom) side to receive the shear wave signal. Finally, the shear wave signal acquired by the receiver is postamplified and then digitized with a Tektronix digital oscilloscope with real-time fast Fourier transform (FFT) capability to produce the transmission power spectrum. For the measurement of layer-TIs, the transmission can be also obtained by changing the layer position of the emitting and receiving ends as detailed in Supplemental Note 6.

## Data availability

The data that support the findings within this work and other related findings are available from the corresponding authors upon reasonable request.

## Author contributions

J.-J.C., Z.-H.T. initiated the program, oversaw and directed the whole project. J.-J.C. and S.-Y.H. conceived the original idea. S.-Y.H. performed the simulations and derived the theory. J.-J.C., S.-Y.H. H.-B.H., and L.-Y.F. supported the fabrication process of the sample. Z.-H.T., S.-Y.H., J.-Y.W., L.-Y.F. and X.-P.X. helped in the experiment setup. Z.-H.T. and J.-J.C. carried out experimental measurements. J.-J.C., S.-Y.H. and H.-B.H. analyzed the data, prepared the figures, and wrote the manuscript. All authors contributed to scientific discussion of the manuscript.

## Acknowledgements

The authors gratefully acknowledge financial support from National Science Foundation of China under Grant No.11374093, No.11672214 and Young Scholar fund sponsored by common university and college of the province in Hunan.

## Additional information

Supplementary information is available in the online version of the paper. Correspondence and requests for materials should be addressed to J.-J.C. or to Z.-H.T. or to X.H.

**Competing financial interests**

The authors declare no competing financial interests.